\documentclass[prd,showpacs,showkeys,floatfix,nofootinbib,fleqn,
               preprint,12pt,tightenlines]{revtex4}

\usepackage{amsmath,amssymb,graphicx,dcolumn}

\newcommand  {\version}{v6}  

\newcommand{\vecbf}[1]{\boldsymbol{\mathrm{#1}}}
\begin{document}
\noindent
Phys. Rev. D 82, 105024 (2010)
\hfill
arXiv:1008.1967 \,(\version)  
\newline\vspace*{0.5cm}
\title[]
      {Potential sensitivities to Lorentz violation from nonbirefringent
modified Maxwell theory of Auger, HESS, and CTA\vspace*{0.25cm}}
\author{F.R.~Klinkhamer}
\email{frans.klinkhamer@kit.edu}
\affiliation{\mbox{Institute for Theoretical Physics, University of Karlsruhe,}\\
             Karlsruhe Institute of Technology, 76128 Karlsruhe, Germany\\}
\begin{abstract}
\vspace*{0.25cm}\noindent
Present and future ultra-high-energy-cosmic-ray facilities
(e.g., the South and North components of the Pierre Auger Observatory)
and TeV-gamma-ray telescope arrays (e.g., HESS and CTA)
have the potential to set stringent bounds on the nine Lorentz-violating
parameters of nonbirefringent modified Maxwell theory
minimally coupled to standard Dirac theory.
A concrete example is given how to obtain, in the coming decennia,
two-sided bounds on the eight anisotropic parameters at the $10^{-20}$
level and an upper (lower) bound on the single isotropic parameter
at the $+10^{-20}$ ($-10^{-16}$) level.
Comparison is made with existing and potential direct bounds
from laboratory experiments.
\end{abstract}

\pacs{11.30.Cp, 12.20.-m, 98.70.Rz, 98.70.Sa}
\keywords{Lorentz violation, quantum electrodynamics, gamma-ray sources,
          cosmic rays}
\maketitle

\section{Introduction}
\label{sec:Introduction}

Nine real dimensionless parameters characterize
nonbirefringent modified Maxwell theory
minimally coupled to the standard Dirac
theory of massive spin--$\textstyle{\frac{1}{2}}$
particles~\cite{ChadhaNielsen1983,ColladayKostelecky1998,KosteleckyMewes2002}.
Recently, new direct laboratory bounds on these parameters have been
published~\cite{Eisele-etal2009,Herrmann-etal2009,Hohensee-etal2010},
specifically, upper bounds on the absolute values of the isolated
nine parameters.
The direct laboratory bounds for the five parity-even anisotropic
parameters are now only 1 order of magnitude above the previous
indirect bounds~\cite{KlinkhamerRisse2008a,KlinkhamerRisse2008b}.
The direct laboratory bounds for the three parity-odd anisotropic
parameters are still 5 orders of magnitude above
the indirect bounds and the situation is even
more dramatic for the single isotropic
parameter~\cite{KlinkhamerSchreck2008}.
(The qualifications `direct' and `indirect' will be explained later.)

With the laboratory bounds steadily
improving~\cite{Herrmann-etal2009,Muller2003Benmessai2008},
we are, therefore, motivated to see what the chances are to
tighten our indirect bounds in the coming decades.
In particular, we will focus on the present and future
capabilities of the Pierre Auger Observatory (PAO)~\cite{Auger2004},
the High Energy Stereoscopic System (HESS) of imaging atmospheric
Cherenkov telescopes~\cite{HESS2003},
and the Cherenkov Telescope Array (CTA) project~\cite{CTA2007-2009}.
As regards gamma-ray telescopes, our focus is on HESS and CTA,
rather than MAGIC~\cite{MAGIC} and VERITAS~\cite{VERITAS},
because the first two offer the best prospects for detecting
photons with the highest possible energies, which will turn out
to be the crucial ingredient for one of our potential bounds.
In that respect, also future extensive-air-shower arrays
look promising~\cite{EAS}.\footnote{An entirely different class
of ``cosmic messengers'' consists of high-energy
neutrinos~\cite{EPNT2006,IceCube2010review,Abbasi-etal2010}, which are
particularly well suited for the search of new Lorentz-violating
and possibly CPT-breaking effects that lie outside the mass
sector~\cite{KlinkhamerVolovik2004,Klinkhamer2006ijmpa,Klinkhamer2006prd}.}

Returning to modified Maxwell theory, it needs to be mentioned that,
in principle, the dimensionless Lorentz-violating parameters of the
photon sector can be of order $1$ if they arise from a nontrivial
small-scale structure of spacetime~\cite{BernadotteKlinkhamer2007}.
The single isotropic Lorentz-violating parameter
$\widetilde{\kappa}_\text{tr}$, in particular, has a natural
interpretation, if positive,
as the excluded-spacetime-volume fraction of ``defects''
embedded in flat Minkowski spacetime.
It is, thus, of fundamental importance to get as tight bounds
as possible, preferably for all nine dimensionless Lorentz-violating
parameters of the nonbirefringent photon sector but certainly for the
isotropic one.

\section{Threshold conditions}
\label{sec:Threshold-conditions}

An interesting suggestion for obtaining bounds on the
Lorentz-violating (LV)
parameters of a hypothetical nonrelativistic theory
is the following~\cite{Beall1970,ColemanGlashow1997,GagnonMoore2004}:
\newcounter{alphcounter}     
\begin{list}{(\alph{alphcounter})}{\usecounter{alphcounter}}
\item
with modified dispersion relations,  new decay
channels may appear which are absent in the
standard relativistic theory;
\item
this leads to rapid energy loss of
particles with energies above threshold
[the threshold energy $E_\text{th}$ depends on the
LV parameters $\kappa$ of the theory, typically with
$E_\text{th}(\kappa)\to\infty$ for $\kappa\to 0$];
\item
observing these particles implies that they necessarily
have energies at or
below threshold ($E\leq E_\text{th}$), which, in turn,
gives bounds on the LV parameters of the theory.
\end{list}
This suggestion relies, of course, on the proper calculation
of the threshold energies and decay rates. See, e.g.,
Ref.~\cite{Lehnert2003} for a general discussion of physically admissible
modified dispersion relations and
Ref.~\cite{KaufholdKlinkhamer2006} for a general discussion of
Lorentz-violating decay processes, starting from simple scalar models.

Consider, now, the photon ($\gamma$) of nonbirefringent modified Maxwell
theory~\cite{ChadhaNielsen1983,ColladayKostelecky1998,KosteleckyMewes2002}
minimally coupled to standard massive spin--$\textstyle{\frac{1}{2}}$
Dirac particles (e.g., the electron $e$ and proton $p$)
in a particular preferred frame, for definiteness, identified with
the sun-centered celestial equatorial frame.\footnote{More precisely,
the theory considered is the one of the standard model with the
experimentally determined values for the coupling constants
and Higgs vacuum expectation value, to which a single nonbirefringent
Lorentz-violating kinetic term of the $U(1)$ gauge bosons is added, namely,
the $k_{B}$ term of Eq.~(16) in Ref.~\cite{ColladayKostelecky1998} or the
$\widetilde{\kappa}^{(1)}$ term of Eq.~(C2a) in Ref.~\cite{KlinkhamerSchreck2008}.
As remarked in Sec. V of Ref.~\cite{KlinkhamerRisse2008a},
the bounds discussed in that article and this one may also apply to
theories with additional Lorentz violation in the quark and lepton sectors,
but, then, a precise formulation of the bounds becomes rather involved.
For this reason, the discussion here is restricted to the
theory with a single Lorentz-violating kinetic term of the
Abelian gauge field.}
Two Lorentz-violating decay processes, in particular,
have been considered: vacuum-Cherenkov radiation
($p^{+} \to p^{+}\gamma$)~\cite{Altschul2007,KaufholdKlinkhamer2007,KlinkhamerSchreck2008}
and photon decay
($\gamma \to e^{+} e^{-}$)~\cite{KlinkhamerSchreck2008,Beall1970,ColemanGlashow1997}.
We start by discussing the first process, vacuum-Cherenkov radiation
of a proton $p$ or heavy nucleus $N$, both, in first approximation,
considered as charged pointlike Dirac particles.

Assume the detection of
an ultra-high-energy-cosmic-ray (UHECR) primary
with energy $E_\text{\,prim}$, mass $M_\text{\,prim}$, and
flight-direction unit vector $\widehat{\vecbf{q}}_\text{\,prim}$. In terms
of the standard-model-extension parameters~\cite{KosteleckyMewes2002},
the vacuum-Cherenkov process ($N_\text{\,prim}\to N_\text{\,prim}\,\gamma$)
then gives the following threshold
condition~\cite{Altschul2007,KaufholdKlinkhamer2007,KlinkhamerRisse2008a,KlinkhamerRisse2008b}
on the LV parameters of nonbirefringent modified Maxwell theory:
\begin{eqnarray}\label{eq:Cherenkov-condition}
R\big[\,2\,\widetilde{\kappa}_\text{tr}
-\widehat{q}_\text{\,prim}^{\;a}\;
 (\widetilde{\kappa}_{\text{o}+})^{bc}\;
 \epsilon^{abc}
-\widehat{q}_\text{\,prim}^{\;a}\;
 (\widetilde{\kappa}_{\text{e}-})^{ab} \;
 \widehat{q}_\text{\,prim}^{\;b}\big]
\leq
\big(M_\text{\,prim}\,c^2/E_\text{\,prim} \big)^2 ,
\end{eqnarray}
where each repeated Cartesian index
$a$, $b$, and $c$ is summed over 1 to 3,
the quantity $\epsilon^{abc}$ stands for
the totally antisymmetric Levi-Civita symbol
with normalization $\epsilon^{123}=1$,
and the ramp function $R$ is defined by
$R[x] \equiv (x + |x|\,)/2$.
The velocity $c \equiv 299\,792\,458\;\text{m/s}$
on the right-hand side of \eqref{eq:Cherenkov-condition}
corresponds, in the Lorentz-violating theory considered,
to the maximal attainable velocity of the Dirac particles.
The nine LV parameters are collected in the antisymmetric
$3\times 3$ matrix $\widetilde{\kappa}_{\text{o}+}$ (three parameters),
the symmetric traceless $3\times 3$ matrix
$\widetilde{\kappa}_{\text{e}-}$ (five parameters),
and the number $\widetilde{\kappa}_\text{tr}$ (one parameter).

The minimal set of UHECR events needed to obtain
significant bounds from \eqref{eq:Cherenkov-condition}
on all nine LV parameters has six events (labeled $l=1,\, \ldots ,\, 6$)
with approximately equal leverage factors
$\big(E_{l}/M_{l}\,c^2\big)^2$ $\gg$ $1$
and approximately (anti)orthogonal flight
directions $\widehat{\vecbf{q}}_{\, l}$.
In an appropriate Cartesian coordinate system,
the ideal flight directions are given by
$\widehat{\vecbf{q}}_{\, 1,2}=(\pm 1,\,     0,\,     0)$,
$\widehat{\vecbf{q}}_{\, 3,4}=(    0,\, \pm 1,\,     0)$, and
$\widehat{\vecbf{q}}_{\, 5,6}=(    0,\,     0,\, \pm 1)$.
Strictly speaking, three orthogonal events suffice
for the parity-even parameters $\widetilde{\kappa}_{\text{e}-}$
and a single event for the positive isotropic
parameter $\widetilde{\kappa}_\text{tr}$.\footnote{Remark that,
with so few events, the bounds are somewhat weakened by the
appearance of geometrical factors in the function argument
on the left-hand side of \eqref{eq:Cherenkov-condition}.
The geometrical factor of the second (parity-odd) term, for example,
is of order $\cos\pi/4 = 1/\sqrt{2}\approx 0.71$.
See also the penultimate paragraph in Sec. II of
Ref.~\cite{KlinkhamerRisse2008a} for further discussion.}
As emphasized in Ref.~\cite{ColemanGlashow1997} and reiterated
in Refs.~\cite{KlinkhamerRisse2008b,KlinkhamerSchreck2008},
these bounds require as only input the mere existence,
at the top of the Earth's atmosphere, of charged cosmic-ray primaries
with travel lengths of a meter or more.
Hence, these bounds are independent of the distance to the
(astronomical) source.

Next, restrict to the isotropic sector of nonbirefringent modified Maxwell
theory and consider the case of a
small negative parameter $\widetilde{\kappa}_\text{tr}$.
Then, vacuum-Cherenkov radiation is no longer possible,
but another decay process becomes available, namely,
photon decay into an electron-positron pair.
The photon-decay process ($\gamma \to e^{+} e^{-}$)
gives the following threshold
condition~\cite{KlinkhamerSchreck2008,Beall1970,ColemanGlashow1997}
on the isotropic LV parameter of modified Maxwell theory:
\begin{equation}\label{eq:photon-decay-condition}
R\big[\,- \widetilde{\kappa}_\text{tr}\big]
\leq
2\;\big( M_{e}\,c^2/E_{\gamma} \big)^2\,,
\end{equation}
with $E_{\gamma}$  the energy of the primary photon, $M_{e}$
the mass of the electron, and $R$ the ramp function defined
below \eqref{eq:Cherenkov-condition}.

The minimal set of gamma-ray events needed to obtain
a significant bound from \eqref{eq:photon-decay-condition}
contains a single event with leverage
factor $\big(E_{\gamma}/M_{e}\,c^2\big)^2\gg 1$
and an arbitrary flight direction (the considered
Lorentz violation being isotropic).
Again, the only input for this bound is the mere existence,
at the top of the Earth's atmosphere, of a primary photon
with a travel length of a meter or more,
making the bound independent of the distance to the
(astronomical) source.\footnote{If it could be
established that certain specific astronomical sources
emit UHECR protons or multi-TeV photons which definitely
radiate their energy away or decay before they reach the Earth,
then a finite range of nonzero values
of the relevant LV parameters could be obtained,
assuming the LV effects to be responsible for the
observed disappearance of the particles.
Such a range could be expected to have a (mild) dependence on the
source distances, whereas the bounds of the present article
are strictly source-distance independent.}

\section{Prototypical data samples}
\label{sec:Prototypical-data-samples}

Two fiducial data samples are presented in this section,
the first with UHECR events for
application of the vacuum-Cherenkov threshold
condition \eqref{eq:Cherenkov-condition}
and the second with TeV gamma rays for
application of the photon-decay threshold condition
\eqref{eq:photon-decay-condition}.
Our aim is to be as concrete as possible,
but it is clear that the data samples considered are
only examples and can certainly be modified or improved
in the future, as will be discussed further in
Sec.~\ref{subsec:Experimental-issues}.

For the first prototypical data sample,
we consider hybrid UHECR events~\cite{PAO2010} detected by Auger
(short for the Pierre Auger Observatory, PAO).
Recall that a so-called `hybrid' Auger event
refers to an extended air shower which has been
observed simultaneously by the two types of detectors,
the water-Cherenkov surface detectors and the fluorescence
telescopes~\cite{Auger2004}.

Specifically, the following
sample $\Sigma_{N^\prime}$ of $N^\prime \equiv N/2$ hybrid Auger
events is obtained from the two energy bins
below $E=10^{19}\,\text{eV}$ in Figs. 2 and 3 of Ref.~\cite{PAO2010}:
\begin{subequations}\label{eq:UHECR-sample}
\begin{eqnarray}
N/2 &=&  131+96  =  227\,,
\label{eq:UHECR-sample-Nover2}\\[2mm]
E & \sim & 8\;\text{EeV}\,,
\label{eq:UHECR-sample-E}\\[2mm]
<  X_\text{max} >
& \sim & 750 \;\text{g}/\text{cm}^{2}\,,
\label{eq:UHECR-sample-Xmax}\\[2mm]
\text{rms}(X_\text{max}) & \sim & 50 \;\text{g}/\text{cm}^{2}\,,
\label{eq:UHECR-sample-rmsXmax}
\end{eqnarray}
where the reason for writing $N/2$ will become clear shortly
and where $<  X_\text{max} >$ stands for the average
of $X_\text{max}$, the atmospheric depth of the shower
maximum (here, determined by the fluorescence telescopes of Auger).
The relative uncertainty of the energy measurement
\eqref{eq:UHECR-sample-E} is taken
equal to $25\,\%$ (slightly larger than the value of $22\,\%$
quoted in Ref.~\cite{PAO2010}).
Remark that the fluorescence telescopes of Auger~\cite{Auger2004,PAO2010}
partly track the primary particle and its secondaries
along their way in the atmosphere, following them
over a length of several kilometers.

The sample $\Sigma_{N/2}$
from \eqref{eq:UHECR-sample-Nover2}--\eqref{eq:UHECR-sample-rmsXmax}
is, however, not quite isotropic,
as the celestial coverage of Auger--South is not complete. Awaiting
results from Auger--North (or a similar cosmic-ray observatory
on the northern hemisphere), we can proceed as follows.
An artificial sample $\Sigma_{N}$ is constructed by
adding to each event of $\Sigma_{N/2}$ one in the opposite direction:
\begin{eqnarray}
\Sigma_{N} =
\big\{\,\{(E_{n},\,\widehat{q}_{\,n},\,C_{n}),\,
         (E_{n},\,-\widehat{q}_{\,n},\,C_{n})\}\,
\big| \, n=1,\, \ldots,\, N/2\,\big\}\,,
\label{eq:UHECR-sample-SigmaN}
\end{eqnarray}
where $\widehat{q}_{\,n}$ is the flight-direction vector
of event $n$ and
$C_{n}$ contains further characteristics such as $X_{\text{max},\,n}$.
In this way, the sample $\Sigma_{N}$ covers the whole
celestial sphere, even though certain regions may be covered
somewhat more densely than others. [There are, of course,
other ways to make the sample $\Sigma_{N/2}$ more isotropic,
for example, by keeping the same number of events but
randomly flipping the flight-direction vectors.]

The last two assumptions are that the sample
$\Sigma_{N}$ contains $N_{p} \gg 1$ events
with proton primaries
and that the arrival directions of these $N_{p}$ events
are representative of an isotropic distribution.
Specifically, we assume that
\begin{eqnarray}
N_{p} &\geq& 48\,,
\label{eq:UHECR-sample-Np}
\end{eqnarray}
and that the proton arrival directions cover the celestial
sphere completely and more or less homogeneously.
The isotropy condition can be described in words as follows:
to have approximately $N_{p}/2$ events in each hemisphere
(solid angle $2\pi\;\text{sr}$),
approximately $N_{p}/4$ events in each quadrant
(solid angle $\pi\;\text{sr}$),
approximately $N_{p}/8$ events in each octant
(solid angle $\pi/2\;\text{sr}$), etc.
Restricting to octants, one possible mathematical formulation
of the isotropy condition for the proton arrival directions is
\begin{eqnarray}
\forall_{n \in \{ 1,\,2,\, \ldots\,,\, 8\}} &:&
 N_{p}\big( \Delta\Omega_{n} =\pi/2\big) \geq 1 \,,
\label{eq:UHECR-sample-Np-isotropy}
\end{eqnarray}
\end{subequations}
which must hold for any choice of octants $\Delta\Omega_{n}$
that cover the celestial sphere ($\Omega =4\pi$) completely.
Condition \eqref{eq:UHECR-sample-Np-isotropy} is not too
difficult to satisfy with the relatively large total number
of protons from \eqref{eq:UHECR-sample-Np}.
For the data sample $\Sigma_{N}$, conditions
\eqref{eq:UHECR-sample-Np} and \eqref{eq:UHECR-sample-Np-isotropy}
could perhaps result from those
events with $X_\text{max} \gtrsim 825 \;\text{g}/\text{cm}^{2}$,
according to the calculations shown in Fig. 3 of Ref.~\cite{PAO2010}
(see also Table I of Ref.~\cite{KlinkhamerRisse2008a}).

The basic idea, now, for obtaining new bounds from
\eqref{eq:Cherenkov-condition}
and the prototypical data sample \eqref{eq:UHECR-sample}
is that, without knowing precisely
which of the $N=454$ events would correspond to a proton,
the knowledge would suffice that there would be a subset
with $N_{p}\gg 1$ proton events distributed isotropically.
Clearly, this requires the reliable determination of the
proton fraction $P_{p}$ in the sample considered
(with the certainty that $P_{p}>0$) and the information that
the protons in this UHECR sample have an overall
isotropy of the arrival directions (irrespective of possible
small-scale clustering). In short, the challenge for Auger
will be to identify a 10--EeV
UHECR sample with enough isotropic protons in it.
Further discussion will be given in
Sec.~\ref{subsec:Experimental-issues}.

Turning to the second prototypical data sample, recall that
$80$ TeV gamma rays have been detected by HEGRA
from the Crab Nebula~\cite{Aharonian-etal2004} and
$10^2$ TeV gamma rays by HESS
from the shell-type supernova remnant
\mbox{RX J1713.7--3946}~\cite{Aharonian-etal2006,Aharonian-etal2007}.
The relative uncertainty of the energy measurement
for this last extended source equals $20\,\%$
according to Ref.~\cite{Aharonian-etal2006}.
Remark that the Cherenkov light detected on the
ground~\cite{EAS-Cherenkov}
indirectly tracks the primary photon and its secondaries
along their way in the atmosphere, essentially following them
over a length of several kilometers.

Specifically, the second sample is taken to have a
single fiducial photon event with an arbitrary arrival direction,
\begin{subequations}\label{eq:gamma-sample}
\begin{eqnarray}
N_{\gamma} &=& 1\,,
\label{eq:gamma-sample-N}\\[2mm]
E_{\gamma} & \sim & 10^2\;\text{TeV}\,.
\label{eq:gamma-sample-E}
\end{eqnarray}\end{subequations}
Here, the crucial assumption is that at least one
such photon has been detected unambiguously
(or, more realistically, that this photon is one of a larger sample
of photon candidates identified with high probability).

\section{Potential sensitivities}
\label{sec:Potential-sensitivities}
\vspace*{-1mm}

Relying on the proton primaries contained in
the prototypical UHECR sample \eqref{eq:UHECR-sample},
the vacuum-Cherenkov threshold condition \eqref{eq:Cherenkov-condition}
could be used for the numerical values
\begin{equation}\label{eq:fiducial-values-Cherenkov-sample}
E_\text{\,prim} \stackrel{?}{=} (8 \pm 2)\;\text{EeV}\,,\quad
M_\text{\,prim} \stackrel{?}{=} M_{p}= 938\;\text{MeV}\,,
\end{equation}
explicitly showing the one-sigma error in the energy determination
and inserting  question marks to emphasize that the data sample
is still artificial.
Assuming a sufficiently large number $N_{p}$ of proton
primaries, the directions $\widehat{\vecbf{q}}_{\text{\,prim},\,l}$
could be taken optimal, so that the geometrical factors entering
the argument on the left-hand side of \eqref{eq:Cherenkov-condition}
would be close to their maximal values.
The main contribution to the statistical error on the bounds
of the LV parameters would come from the energy uncertainty and
would correspond to approximately $50\,\%$ for this sample.

The resulting two--$\sigma$ indirect bounds on the nine isolated
LV parameters of nonbirefringent modified Maxwell theory would be:
\begin{subequations}\label{eq:improved-SMEbounds}
\begin{eqnarray}
\big| (\widetilde{\kappa}_{\text{o}+})^{(ab)}\,
\big|_{\,(ab)\,=\,(23),\,(31),\,(12)}
&\stackrel{?}{<}& 1.4 \times 10^{-20}\,,
\label{eq:improved-SMEbounds-anisotropic-odd}\\[1mm]
\big| (\widetilde{\kappa}_{\text{e}-})^{(ab)}\,
\big|_{\,(ab)\,=\,(11),\,(12),\,(13),\,(22),\,(23)}
&\stackrel{?}{<}& 3 \times 10^{-20}\,,
\label{eq:improved-SMEbounds-anisotropic-even}\\[1mm]
\widetilde{\kappa}_\text{tr} &\stackrel{?}{<}& 1.4 \times 10^{-20}\,,
\label{eq:improved-SMEbounds-isotropic}
\end{eqnarray}\end{subequations}
which, for the first two entries, would improve by 2 orders of
magnitude and, for the last entry, by 1 order of magnitude
upon the previous bounds in Ref.~\cite{KlinkhamerRisse2008b}.
There would be no mystery where the improvement would come from:
comparing the values \eqref{eq:fiducial-values-Cherenkov-sample}
to those used in Ref.~\cite{KlinkhamerRisse2008b},
the gain from a much smaller mass value $M_\text{\,prim}$
would overcome the loss from a somewhat larger value of the
inverse energy $1/E_\text{\,prim}$.

With the prototypical gamma-ray sample \eqref{eq:gamma-sample},
the photon-decay threshold
condition \eqref{eq:photon-decay-condition} could be used with
numerical values
\begin{equation}\label{eq:fiducial-values-gamma-sample}
E_{\gamma} \stackrel{?}{=} (100 \pm 20)\;\text{TeV}\,,\quad
M_{e} = 0.511\;\text{MeV}\,.
\end{equation}

The resulting two--$\sigma$ indirect lower bound
on the single isotropic LV parameter
of nonbirefringent modified Maxwell theory would be
\begin{equation}\label{eq:improved-HESS-lowerbound}
\widetilde{\kappa}_\text{tr} \stackrel{?}{>} - 0.9 \times 10^{-16}\,,
\end{equation}
which would improve by 1 order of magnitude upon the previous
two--$\sigma$ lower
bound $\widetilde{\kappa}_\text{tr} > - 0.9 \times 10^{-15}$
\cite{KlinkhamerSchreck2008}.\footnote{The qualitative
lower bound on $\widetilde{\kappa}_\text{tr}$ at the $- 10^{-16}$
level (no confidence level specified) from
Ref.~\cite{SteckerGlashow2001}
relies on a detection of $50\;\text{TeV}$ gamma rays from the
Crab Nebula, which is of marginal significance~\cite{Tanimori-etal1998}.}
Again, there would be no mystery where the improvement would
come from, namely, the somewhat larger energy value $E_{\gamma}$
in \eqref{eq:fiducial-values-gamma-sample} compared to
the one used in Ref.~\cite{KlinkhamerSchreck2008}.

Both indirect bounds \eqref{eq:improved-SMEbounds}
and \eqref{eq:improved-HESS-lowerbound} would scale with the
inverse of the energy square of the incoming particle, that is,
with $\big[(8\;\text{EeV})/E_\text{\,prim}\big]^2$ and
$\big[(10^2\;\text{TeV})/E_{\gamma}\big]^2$, respectively.
But, for the moment, these bounds are only potential bounds
as they rely on certain assumptions (e.g., $N_{p} \gg 1$)
or on low-statistics data (e.g., $E_{\gamma} = 10^2\;\text{TeV}$).

\section{Discussion}
\label{sec:Discussion}
\vspace*{-2mm}

\subsection{Experimental issues}
\label{subsec:Experimental-issues}
\vspace*{-2mm}

As mentioned in the previous section,
bounds \eqref{eq:improved-SMEbounds}
and \eqref{eq:improved-HESS-lowerbound} are, for the
moment, only indicative of the potential sensitivity
of Auger and HESS/CTA, because they either rely on certain
assumptions or use low-statistics data. We expect that
both bounds will be realized in the coming decades, derived
without additional assumptions and from high-statistics data.
Let us now discuss the arguments on which
this expectation is based.

The potential UHECR bounds \eqref{eq:improved-SMEbounds}
rely on two main assumptions for the event sample considered:
(i) the large-scale isotropy of the UHECR arrival directions
(applicable to all primary subspecies), and
(ii) the presence of a nonvanishing fraction of proton primaries
or, at least, of light-nucleus primaries.
Both assumptions will be tested (and, most likely, verified)
with further Auger--South data and forthcoming Auger--North data.

At this point, it is already clear that it will be
of critical importance for Auger to obtain, for a subset of events,
a reliable determination of the
type of primary (e.g., a hydrogen, helium, oxygen, or iron nucleus),
preferably by use of different types of diagnostics
(for example, in addition to $ X_\text{max}$,
number densities of muons and arrival times of particles
reaching the surface~\cite{UHECR-composition}).
As mentioned in Sec.~\ref{sec:Threshold-conditions},
only six (anti)orthogonal events with
$(M_\text{\,prim}\,c^2/E_\text{\,prim})^2 \sim 10^{-20}$ would suffice,
as long as $M_\text{\,prim}$ would be determined accurately.

Alternatively, a nonzero fraction $P_{\,i}$ of primary type $i$
(for example, $i=p$, He, O, or Fe)
would suffice for establishing the same bound, provided
a large enough sample of isotropic events would be available
with $(M_{\,i}\,c^2/E_{\,i})^2 \sim 10^{-20}$
[this is, in fact, the strategy followed with the
prototypical data sample \eqref{eq:UHECR-sample}
used to obtain the potential bounds \eqref{eq:improved-SMEbounds}
for $i=p$ and $M_{\,i} = 0.938\;\text{GeV}$].
This strategy, with a nonzero $P_{\,i}$ for an appropriate
sample of UHECR events,
appears the most promising for improving the
bounds of Ref.~\cite{KlinkhamerRisse2008b}.

Recent observations from Auger~\cite{PAO2010}
indicate an increasing fraction of heavy nuclei with
increasing primary energy $E$. Still, it is very well possible that
the fraction of light nuclei (e.g., hydrogen) is small
but nonzero at any fixed value of $E$.
In fact, a recent suggestion~\cite{Calvez-etal2010}
to explain the rising fraction of ironlike nuclei
seems to imply just that (e.g.,
$P_{\,p}\sim 10\,\%$ and $P_\text{\,Fe}\sim 90\,\%$
at $E \sim 8\,\text{EeV}$ for the simple model of
Fig.~1 of Ref.~\cite{Calvez-etal2010}).
It, then, remains to determine the optimal primary type $i$
with the smallest value of $M_{\,i}\,c^2/E_{\,i}$ and
enough events ($P_{\,i}\,N \gtrsim 10$) having arrival directions
more or less isotropically distributed
over the celestial sphere (irrespective of the
positions of the  original UHECR sources).
As mentioned already in Sec.~\ref{sec:Prototypical-data-samples},
the argument for obtaining the bound is
statistical,  without the need
to know precisely which event of the $N$--event sample corresponds to
species type $i$, as long as we can be sure about the nonzero
value of the fraction $P_{\,i}$ and the arrival-direction isotropy.

Now turn to the potential gamma-ray
bound \eqref{eq:improved-HESS-lowerbound},
which relies on the detection of $10^2\;\text{TeV}$
gamma rays by HESS. For the moment, this detection has
only a marginal significance (less than two--$\sigma$
according to Fig.~3 and Table~5 of Ref.~\cite{Aharonian-etal2006}).
Future improvement can be expected from further HESS
(or VERITAS) observations and
certainly with CTA~\cite{CTA2007-2009}, which may have a 10 times
greater sensitivity than HESS at energies $E_{\gamma} \sim 10^2\;\text{TeV}$.
If CTA (or an extensive-air-shower array~\cite{EAS})
unambiguously detects $10^2\;\text{TeV}$ photons
from whichever astronomical source, the potential
bound \eqref{eq:improved-HESS-lowerbound} will be transformed
into one of the cleanest bounds on isotropic Lorentz violation
in the photon sector.

In addition, there is always the possibility of a detection by Auger
of ultra-high-energy photons~\cite{RisseHomola2007,PAO-photon2009}.
Note that a photon event with $E_{\gamma} = 10^{18}\;\text{eV}$,
for example,
would improve the lower bound \eqref{eq:improved-HESS-lowerbound},
based on \eqref{eq:photon-decay-condition} and
\eqref{eq:fiducial-values-gamma-sample},
by a factor $10^{8}$ to the level of $- 0.9 \times 10^{-24}$.

\subsection{Theoretical issues}
\label{subsec:Theoretical-issues}

Following up on these experimental considerations,
we have two remarks on theoretical issues,
one regarding the distinction of `direct' vs. `indirect' bounds already
made in Sec.~\ref{sec:Introduction}
and the other regarding the distinction of `earth-based' vs.
`astrophysical' bounds.

First, a direct bound on the LV parameters of nonbirefringent
modified Maxwell theory~\cite{ChadhaNielsen1983,ColladayKostelecky1998,KosteleckyMewes2002}
is taken to refer to experiments which directly
test the modified propagation properties of the nonstandard photons,
for example, by measuring positions and arrival times of pulses of light.
An indirect bound, instead, is taken to refer
to experiments which look for derived
effects resulting from the nonstandard dispersion relations
of the photons. These derived effects
[here, vacuum-Cherenkov radiation and photon decay]
are, however, unambiguously calculable from the
theory considered in this article, which has
a modified pure-photon sector and a
standard matter sector with minimal photon-matter couplings
(governed by the principle of gauge invariance).
In fact, the energy thresholds used for the indirect bounds
of Refs.~\cite{KlinkhamerRisse2008b,KlinkhamerSchreck2008}
and the present article only rely on energy-momentum conservation.
For this reason, we do not consider indirect bounds
to be necessarily less reliable than direct bounds.

Second, it is important to realize
that the existing indirect bounds from
Refs.~\cite{KlinkhamerRisse2008b,KlinkhamerSchreck2008}
and the potential indirect bounds from the present article are earth-based
bounds (the atmosphere of the Earth corresponding to the `front end'
of the detector~\cite{Auger2004,HESS2003}) and not astrophysical
bounds, which may rely on additional assumptions about the astronomical
source and the propagation distance of the primary particles.
We have already mentioned in Secs.~\ref{sec:Threshold-conditions}
and \ref{sec:Prototypical-data-samples} that our indirect LV
bounds require the existence of primaries with track lengths
of a meter and that Auger measures shower tracks
which start at a point high up in the Earth's atmosphere
and run for several kilometers. Hence, we can simply
repeat what was written in Ref.~\cite{KlinkhamerRisse2008b},
``these Cherenkov bounds only depend on the measured energies
and flight directions of the charged cosmic-ray primaries
at the top of the Earth atmosphere.'' The same
holds for the photon-decay bound, now referring to
measurements of the photon primary. As such, these earth-based bounds
are not inherently inferior to laboratory bounds.

\subsection{Comparison of bounds}
\label{subsec:Comparison-of-bounds}

Table~\ref{tab:bounds-compared} now compares the different types
of bounds for the parameters of nonbirefringent modified Maxwell
theory minimally coupled to standard Dirac theory.
The second data row of this table refers to a potential
sensitivity of order $10^{-20}$~\cite{Herrmann-etal2009}
which may be achieved with cryogenic
resonators~\cite{Muller2003Benmessai2008}.
As it stands, direct laboratory experiments are affected
by the earth-velocity boost factor $\beta_{\bigoplus} \sim 10^{-4}$,
which reduces the sensitivity for the parity-odd parameters
$\widetilde{\kappa}_{\text{o}+}$ linearly and for
the isotropic parameter $\widetilde{\kappa}_\text{tr}$
quadratically. There has been a suggestion~\cite{MewesPetroff2006}
of how, in principle, this  boost-dependence problem
may be overcome for the parity-odd parameters,
but, to the best of our knowledge,
this has not yet been realized experimentally.

As the effects of the Earth's velocity are negligible for
the indirect bounds from vacuum-Cherenkov radiation and
photon decay~\cite{KlinkhamerRisse2008a,KlinkhamerRisse2008b,KlinkhamerSchreck2008},
the corresponding indirect bounds for the parity-odd
and isotropic parameters are much stronger than the direct
bounds.\footnote{The same holds for an indirect laboratory bound
(relying on derived effects for the synchrotron radiation rate at LEP)
which gives $|\widetilde{\kappa}_\text{tr}|$ $<$
$5 \times 10^{-15}$ at the two--$\sigma$ level~\cite{Altschul2009}.}
Note also that it is entirely possible that future
data samples from Auger--South+North and CTA
may provide us with 3 times larger energies
than used in \eqref{eq:fiducial-values-Cherenkov-sample}
and \eqref{eq:fiducial-values-gamma-sample},
so that the sensitivities of the last data row
in Table~\ref{tab:bounds-compared}
would be improved by an order of magnitude to the level of $10^{-21}$
for the eight anisotropic
parameters\footnote{Inferred
two--$\sigma$ bounds on the eight anisotropic parameters
already exist at the $10^{-20}$ level
(even better for the parity-even parameters) from an
atomic-fountain-clock laboratory experiment~\cite{Wolf-etal2006},
if a coordinate transformation is used to move the isolated
Lorentz violation in the proton sector (eight parameters)
to the electromagnetic sector (eight parameters).
For further discussion of these coordinate transformations,
see, e.g., App. A in Ref.~\cite{BaileyKostelecky2004}
and Sec.~VII in Ref.~\cite{LammerzahlHehl2004}.}
and $(-10^{-17},\,+10^{-21})$ for the single isotropic parameter.
Especially, the latter projected CTA/Auger sensitivities
for the isotropic parameter $\widetilde{\kappa}_\text{tr}$
would be a welcome complement to future laboratory bounds.

\begin{table*}[t]
\vspace*{0mm}
\begin{center}
\caption{Orders of magnitude for existing and potential
two--$\sigma$ bounds on the nine Lorentz-violating
parameters of nonbirefringent modified Maxwell
theory~\cite{ChadhaNielsen1983,ColladayKostelecky1998,KosteleckyMewes2002}
(minimally coupled to standard Dirac theory)
from direct laboratory and indirect earth-based experiments.
The considered indirect bounds for the isotropic parameter
$\widetilde{\kappa}_\text{tr}$ have different orders of magnitude
depending on the sign of $\widetilde{\kappa}_\text{tr}$
[the values of the last data row, for example, correspond
to \eqref{eq:improved-HESS-lowerbound} and
\eqref{eq:improved-SMEbounds-isotropic} in the main text].
The second data row for potential direct laboratory bounds shows
explicitly the effects from
the earth-velocity boost factor $\beta_{\bigoplus} \sim 10^{-4}$.%
\vspace*{5mm}} \label{tab-highE-Auger-events}
\renewcommand{\tabcolsep}{1.1pc}    
\renewcommand{\arraystretch}{1.1}   
\begin{tabular}{lcccc}
\hline\hline
  Type of bound
& $\widetilde{\kappa}_{\text{e}-}$
& $\widetilde{\kappa}_{\text{o}+}$
& $\widetilde{\kappa}_\text{tr}$
& Reference\\
\hline
  Existing, direct
& $10^{-17}$
& $10^{-13}$
& $10^{-8}$
& \cite{Eisele-etal2009,Herrmann-etal2009,Hohensee-etal2010} \\
Potential, direct
& $10^{-20}  \,?$
& $10^{-20+4}\,?$
& $10^{-20+8}\,?$
&\cite{Herrmann-etal2009,Muller2003Benmessai2008}\\
Existing, indirect
& $10^{-18}$
& $10^{-18}$
& $\big(-10^{-15}\,,\,+10^{-19}\,\big)$
& \cite{KlinkhamerRisse2008b,KlinkhamerSchreck2008}\\
Potential, indirect
& $10^{-20}\,?$
& $10^{-20}\,?$
& $\big(-10^{-16}\,?\,,\,+10^{-20}\,?\,\big)$
& This paper \\
\hline\hline
\label{tab:bounds-compared}
\end{tabular}
\end{center}
\end{table*}

\section{Conclusion}
\label{sec:Conclusion}

The current direct laboratory
bounds~\cite{Eisele-etal2009,Herrmann-etal2009,Hohensee-etal2010}
on the nine Lorentz-violating parameters of nonbirefringent
modified Maxwell
theory~\cite{ChadhaNielsen1983,ColladayKostelecky1998,KosteleckyMewes2002}
minimally coupled to standard Dirac theory are absolutely remarkable.
Still, they fall short of what has been achieved
indirectly~\cite{KlinkhamerRisse2008b,KlinkhamerSchreck2008},
as can be seen by comparing the first and third
data rows in Table~\ref{tab:bounds-compared}.
Instead of arguing about the relative merits of these direct and
indirect bounds (cf. Sec.~\ref{subsec:Theoretical-issues}), we adopt
the following pragmatic approach.

The existing indirect earth-based bounds of
Refs.~\cite{KlinkhamerRisse2008b,KlinkhamerSchreck2008}
can be used to predict that future direct or indirect
laboratory experiments will not
detect nonbirefringent Lorentz violation in the photon sector
at levels of order $10^{-18}$ for the eight anisotropic parameters
and $10^{-15}$ for the single isotropic parameter.
Furthermore, this article has shown that
experiments at present and future observatories
such as Auger and CTA have the potential to reduce these
predictions to the $10^{-20}$ level for the anisotropic
parameters and to the $10^{-16}$ level for the isotropic
parameter (see the last data row in Table~\ref{tab:bounds-compared}).
Perhaps the most interesting predictions of Auger and CTA would
be for the isotropic parameter $\widetilde{\kappa}_\text{tr}$,
which might ultimately drop to levels of order
$(-10^{-17},\,+10^{-21})$ for energy values
$E_\text{\,prim} = 25\;\text{EeV}$ and
$E_{\gamma}      = 3\times 10^{2}\;\text{TeV}$
replacing those of
Eqs.~\eqref{eq:fiducial-values-Cherenkov-sample}
and \eqref{eq:fiducial-values-gamma-sample}, respectively.

\section*{\hspace*{-5.5mm}ACKNOWLEDGMENTS}
\noindent
It is a pleasure to thank W.~Hofmann, R. Lehnert, and M.~Risse
for helpful discussions.


\end{document}